# Time-resolved thermal lens spectroscopy of glassy dynamics in supercooled liquids: theory and experiments

Pengfei Zhang[1], Marco Gandolfi[1,2,3,4,5], Francesco Banfi[5,6], Christ Glorieux[1], and Liwang Liu[1*]

[1]*Laboratory for Soft Matter and Biophysics, Department of Physics and Astronomy, KU Leuven, Celestijnenlaan 200D, B-3001 Heverlee, Belgium*
[2]*CNR-INO, Via Branze 45, 25123 Brescia, Italy*
[3]*Department of Information Engineering, University of Brescia, Via Branze 38, 25123 Brescia, Italy*
[4]*Dipartimento di Matematica e Fisica, Università Cattolica del Sacro Cuore, Via Musei 41, 25121 Brescia, Italy*
[5]*Interdisciplinary Laboratories for Advanced Materials Physics (I-LAMP), Via Musei 41, 25121 Brescia, Italy*
[6]*FemtoNanoOptics group, Université de Lyon, CNRS, Université Claude Bernard Lyon 1, Institut Lumière Matière, F-69622 Villeurbanne, France*

## ABSTRACT

Specific heat and linear thermal expansivity are fundamental thermal dynamics and have been proven as interesting relaxing quantities to investigate in glass transition and glassy state. However, their possibility has much less been exploited compared to mechanical and dielectric susceptibilities due to the limited spectroscopy bandwidth. This work reports on simultaneous spectroscopy of the two by making use of ultrafast time-resolved thermal lens (TL) spectroscopy. Detailed modeling of the thermoelastic transients of a relaxing system subjected to ultrashort laser heating is presented to describe the TL response. The model has been applied to analyze a set of experimentally recorded TL waveforms, allowing the determination of relaxation strength and relaxation frequency from sub-kilohertz to sub-100 MHz and in a wide temperature range from 200-280 K.

## 1. INTRODUCTION

Upon cooling below the freezing point, most liquids can avoid crystallization [1,2], if the cooling takes place sufficiently fast, and arrive at a metastable glassy state. This phenomenon, i.e., supercooling or undercooling, has gained tremendous research interest for decades [3] owing to its impact on a various branch of science and technology, e.g., energy storage [4], food manufacturing [5], and pharmaceuticals development [6]. Supercooled materials exhibit a frequency-dependent response to different kinds of stimuli [7–9], with a strongly temperature-dependent relaxation frequency/time [10,11], and with a moderately temperature-dependent relaxation strength [10]. The most direct presentation of the feature is expressed in the elastic moduli of a relaxing system, which behave more rigidly upon faster stimulation due to a decreasing amount of possibilities for cooperative molecular motions within an oscillation period. The characteristic relaxation frequency, below which the system behaves substantially more softly than in the high frequency limit, drastically decreases with decreasing temperature. How rapidly the relaxation becomes sluggish towards glass transition is defined as the so-called fragility of a system, a signature of the slowing-down process and reflecting to what extent the activation energy decreases with increasing temperature in a potential energy landscape model [12]. As the temperature dependence of relaxation is strong, many efforts have been made to develop and combine different experimental approaches to address relaxation dynamics in a broad frequency range. Dielectric and mechanical spectroscopy [13] is the most frequently used techniques owing to their extraordinary bandwidth, covering respectively 18 decades [14] and 13 decades [15] and enabling the test of several key predictions and models developed in glass physics in a wide temperature range, e.g., time-temperature superposition [16,17], power-law [18], or mode coupling theory [19–21]. Along with dielectric and mechanical relaxation, thermal relaxation [17,22–24] has also proven to be valuable to investigate as it is closely tied with the thermodynamics of the system and couples all degree of freedom equally, which is not true for dielectric or mechanical susceptibilities [25,26]. In practice, thermal relaxation has been observed as a frequency dependence of the specific heat capacity $C(\omega)$, through the 3-$\omega$ technique [26,27], photopyroelectric spectroscopy (PPE) [28,29], and AC-chip nanocalorimetry [30], and of the thermal expansion coefficient $\gamma(\omega)$, through capacitive scanning dilatometry [31–33]. However, till now, the possibility of thermal relaxation has been much less exploited than mechanical and dielectric relaxation because of the frequency range that could be experimentally covered by thermal response techniques being too narrow, about 100 kHz for specific heat spectroscopy [28,29] and only 1 Hz for the thermal expansivity spectroscopy [31–33]. Interestingly, it has been observed in some glassformers that the rotational motion slows down more dramatically than the translational motion [34,35], namely the so-called time-scale decoupling [17,36]. The thermal relaxation couples all the motion and weights all degrees of freedom in the liquid equally, which raises the question whether the decoupling appears also in the thermal relaxation dynamics. Extending the bandwidth to higher frequencies is crucial in order to make the thermal susceptibilities more useful relaxing quantities to investigate, and importantly, to enable the possibility to compare over a substantial wide frequency range the relaxation of the specific heat (all motional degrees of freedom) with one of the dielectric permittivity (rotational mobility) and elastic moduli (translational mobility), and hence to address actual questions on the universality of the fragility value between different relaxing quantities.

---

* liwang.liu@kuleuven.be

In this work, we report the broadband spectroscopy of $C(\omega)$ and $\gamma(\omega)$ till sub-100 MHz based on the use of ultrafast time-resolved Thermal Lens (TL) spectroscopy, in which the transient density response of a relaxing sample subjected to the ultrashort pulse laser heating is exploited to investigate the structural relaxation behavior. Detailed theoretical modeling of the time-resolved TL signals in a relaxing system is presented. An experimental TL spectroscopy of the thermal relaxation dynamics, $C(\omega)$ and $\gamma(\omega)$, in supercooled glycerol is illustrated. Key relaxation features, e.g., low/high-frequency limit response, relaxation strength, and characteristic frequencies are determined and compared with those of mechanical and dielectric relaxation, determined by other techniques.

The paper is structured as follows: In Section 2, we present in detail the modeling of TL response in a glassy system, starting from the calculation of the temperature response to the impulse photothermal excitation, to the derivation of thermoelastic coupling in a relaxing system, and finally an analytical expression to describe the time-resolved TL response based on Fourier optics. Section 3 shows an experimental TL spectroscopy of the slowing down glassy dynamics in supercooled glycerol. The model developed in this work has been used to extract the $C(\omega)$ and $\gamma(\omega)$ in a board frequency range, from sub-kilohertz to tens of MHz, and in a wide temperature range from 200-280 K. Conclusions and perspectives of the approach are given in Section 4.

## 2. THEORETICAL MODELING OF TL RESPONSE IN RELAXING SYSTEMS

TL spectroscopy is a photothermal method that detects the temperature variation in a sample due to heat generated from non-radiative relaxation processes resulting from optical absorption of light. It has been widely used for thermo-optical characterization of materials, spectrometry of photochemical reactions, microvolume and trace analyses of gas and liquids [37,38]. This work extends its application towards spectroscopy of glassy dynamics, especially the relaxation of specific heat and thermal expansivity. We start with the theoretical modeling of time-resolved TL response in a relaxing system.

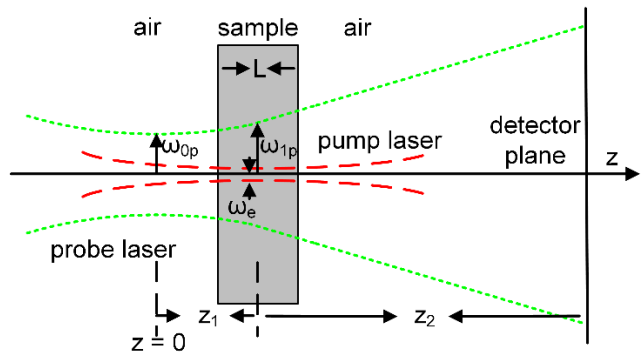

FIG. 1   A schematic diagram of the geometric position of the beams in a TL experiment. The position of the waist of the probe beam is taken as the origin ($z = 0$) along the axis $z$. The sample is located at the distance $z_1$ from the origin. A focused pump beam of spot size $\omega_e$ is used to photothermally excite the sample. A probe beam of the size of $\omega_{0p}$ is used to detect the TL response, which is recorded at the detection plane located at the distance of

$z_2$ from the sample center.

Fig. 1 shows a typical beam geometry in TL experiments, in which coaxially aligned pump (red) and probe (green) laser beams are focused into the bulk of a weakly absorbing sample that is sealed in a cuvette. Local photothermal heating near the pump beam waist produces a transverse temperature gradient and gives rise to a refractive index gradient, hence a TL (gradient-index). The TL may behave like a concave or a convex lens, depending on the thermo-optical coefficient of the sample, which perturbs the wavefront of the propagating probe beam, e.g., beam divergence. The calculation of the TL impulse response of the density of the sample can be divided into two parts: the first calculates the temperature distribution in the sample and the second derives the density response caused by the thermally induced radial strain, namely a combination of the thermal strain and acoustic strain.

## 2.1 Photothermally induced temperature field in a relaxing system

Given a weakly absorbing system, in a TL experiment scheme, the heat distribution along the beam direction or $z$-direction can be considered uniform since the depth of focus of the pump beam is much smaller than optical penetration depth. The temperature field then depends only on the radial direction. The heat diffusion equation can be written in cylindrical coordinates as [39],

$$\frac{1}{r}\frac{\partial}{\partial r}(r\frac{\partial \Delta T(r,t)}{\partial r}) - \frac{1}{\alpha}\frac{\partial \Delta T(r,t)}{\partial t} = -\frac{Q(r,t)}{\kappa} \quad (1)$$

with the boundary condition as $\Delta T(r = R, t) = 0$, $R$ [mm] is the radius of sample. $\alpha = \kappa/\rho C$ [m$^2$ s$^{-1}$] is the thermal diffusivity, $\kappa$ [W m$^{-1}$ K$^{-1}$], $\rho$ [kg m$^{-3}$] and $C$ [J kg$^{-1}$ K$^{-1}$] are the thermal conductivity, density, and specific heat capacity, respectively. $Q(r, t)$ [J s$^{-1}$ m$^{-3}$] is the absorbed heat power density, which in the case of TEM00 Gaussian excitation beam can be expressed in paraxial approximation as [39],

$$Q(r,t) = Q_0 \exp(-\frac{2r^2}{\omega_e^2})\delta(t) \quad (2)$$

with $Q_0$ [J m$^{-3}$] the supplied heat density. The Dirac delta function in time $\delta(t)$ [s$^{-1}$] represents the pulsed excitation. The physical meaning of this condition is that the system is initially at equilibrium, i.e., $\Delta T(r, t) = 0$ for $t < 0$ and at time $t = 0$ an ultrashort pulse excites the system. After Fourier transform, the temperature distribution in the frequency domain can be written as,

$$\frac{1}{r}\frac{\partial}{\partial r}(r\frac{\partial \Delta \tilde{T}(r,\omega)}{\partial r}) - \frac{i\omega}{\alpha}\Delta \tilde{T}(r,\omega) = -\frac{Q_0}{2\pi\kappa}\exp(-\frac{2r^2}{\omega_e^2}) \quad (3)$$

with $i = \sqrt{-1}$. It is known that functions defined on a finite interval can be expanded in terms of a Fourier Bessel series [40]. Thus, the initial solution of Eq. 3 can be written as,

$$\Delta \tilde{T}(r,\omega) = \sum_{n=1}^{\infty} J_0(q_n r)\theta_n(\omega) \tag{4}$$

where $q_n = j_{0n}/R$, $j_{0n}$ denotes the $n^{\text{th}}$ root of the zero-order Bessel function of the first kind $J_0(x)$, and $\theta_n(\omega)$ is the corresponding Fourier Bessel coefficients. Inserting Eq. 4 into Eq. 3 (see Appendix A for detailed calculation), one can get,

$$\sum_{n=1}^{\infty} J_0(q_n r)\theta_n(\omega)\left(\kappa q_n^2 + i\omega\rho C\right) = \frac{Q_0}{2\pi}\exp(-\frac{2r^2}{\omega_e^2}) \tag{5}$$

Then, the unknown Fourier Bessel coefficients $\theta_n(\omega)$ can be determined by using the orthogonality properties of the Bessel function [41]. Multiplying $rJ_0(q_n r)$ in both sides of Eq. 5 and integrating over $r$ from 0 to $R$, we can get,

$$\theta_n(\omega) = \frac{Q_0 I_n}{\pi J_1^2(j_{0n})R^2} \times \frac{1}{\kappa q_n^2 + i\omega\rho C} \tag{6}$$

where $J_1(x)$ is the 1$^{\text{st}}$ order Bessel function of the first kind. $I_n$ can be approximated as,

$$I_n = \int_0^R \exp(-\frac{2r^2}{\omega_e^2})J_0(q_n r)rdr \approx \int_0^{\infty} \exp(-\frac{2r^2}{\omega_e^2})J_0(q_n r)rdr = \frac{\omega_e^2}{4}\exp(-\frac{\omega_e^2 q_n^2}{8}) \tag{7}$$

This approximation is reasonable as the radius of the sample is much bigger than the excitation beam waist ($R$ = 30 mm, $\omega_e \approx$ 30 μm in our experiment). Furthermore, the last passage is justified using the Hankel transform of the Gaussian function. In the impulsive stimulated TL experiment of glassy systems, complex frequency-dependent behavior is found in the specific heat capacity, following the Debye relaxation model,

$$C(\omega) = C_{\infty} + \frac{\Delta C}{1 + i\omega/\omega_C} \tag{8}$$

with $\Delta C = C_0 - C_{\infty}$, $C_{\infty}$ the high-frequency limit and $C_0$ the low-frequency (or static) limit of specific heat capacity. $\omega_C$ [s$^{-1}$] is the characteristic relaxation (angular) frequency, strongly depending on temperature. Thus, the final temperature distribution in the frequency domain can be obtained by combining the aforementioned relations,

$$\Delta \tilde{T}(r,\omega) = \sum_{n=1}^{\infty} J_0(q_n r)\frac{Q_0 I_n}{\pi\rho C_{\infty} J_1^2(j_{0n})R^2} \times \frac{-i(\omega - i\omega_C)}{(\omega - \omega_{1n})(\omega - \omega_{2n})} \tag{9}$$

with,

$$\begin{cases} \omega_{1n} = \dfrac{i}{2}\left(q_n^2\alpha_\infty + \omega_C(1+\dfrac{\Delta C}{C_\infty}) - \sqrt{-4q_n^2\alpha_\infty\omega_C + (q_n^2\alpha_\infty + \omega_C(1+\dfrac{\Delta C}{C_\infty}))^2}\right) \\ \omega_{2n} = \dfrac{i}{2}\left(q_n^2\alpha_\infty + \omega_C(1+\dfrac{\Delta C}{C_\infty}) + \sqrt{-4q_n^2\alpha_\infty\omega_C + (q_n^2\alpha_\infty + \omega_C(1+\dfrac{\Delta C}{C_\infty}))^2}\right) \end{cases} \quad (10)$$

where $\alpha_\infty = \kappa/\rho C_\infty$. The advantage of the notation in Eq. 9 is that it is easy to transform back to the time domain by making use of the residue theorem (see Appendix B for detailed expression) and write $\Delta T(r, t)$ as a combination of two exponentially damped contributions,

$$\Delta T(r,t) = \sum_{n=1}^{\infty} J_0(q_n r) \frac{2Q_0 I_n}{\rho C_\infty J_1^2(j_{0n})R^2} \times \left(\frac{(\omega_{1n}-i\omega_C)}{\omega_{1n}-\omega_{2n}}\exp(i\omega_{1n}t) + \frac{(\omega_{2n}-i\omega_C)}{\omega_{2n}-\omega_{1n}}\exp(i\omega_{2n}t)\right)H(t) \quad (11)$$

where $H(t)$ is the Heaviside step function. If specific heat capacity is independent of frequency (i.e., $\Delta C = 0$), $\omega_{1n}$ and $\omega_{2n}$ reduce to $i\omega_C$ and $iq_n^2\alpha_0$, respectively, where $\alpha_0 = \kappa/\rho C_0$. Then the expression for temperature will reduce to the well-known thermal diffusion equation with $q_n^2\alpha_0$ as the thermal diffusion coefficient,

$$\Delta T(r,t) = \sum_{n=1}^{\infty} J_0(q_n r) \frac{2Q_0 I_n}{\rho C_\infty J_1^2(j_{0n})R^2} \times \exp(-q_n^2\alpha_0 t)H(t) \quad (12)$$

Note that in absence of the diffusion term or by setting $t = 0$ in Eq. 12, the Bessel series corresponds with the Bessel expansion of the spatial part of the heat source. In other words, at $t = 0$, the temperature profile corresponds with the heat source profile. In the presence of thermal diffusion, every Bessel wavenumber component diffuses exponentially with thermal diffusion time $1/q_n^2\alpha_0$.

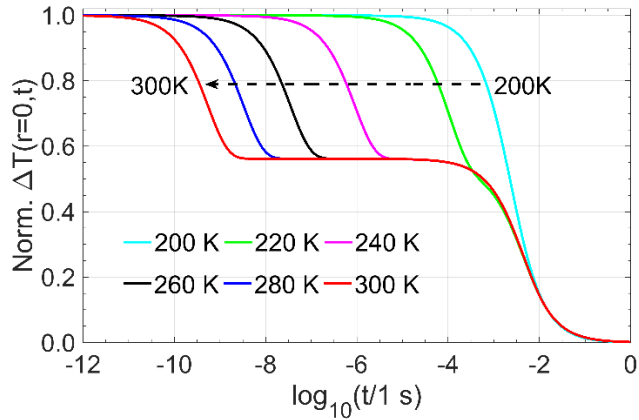

FIG. 2   Simulated normalized temperature response at the focal point of the pump beam ($r = 0$), with input parameters from glycerol summarized in Table 1, plotted on a $\log_{10}$ timescale. The normalization was done by dividing the temperature response to its maximum.

As the specific heat capacity is frequency dependent, the thermal diffusion coefficient $q_n^2\alpha_0$ splits into two values. An extra exponential term in Eq. 11 due to the frequency dependence will cause a temperature dependent overshoot in the temperature response. Assuming characteristic relaxation frequency $\omega_C$ follows Vogel–Fulcher–Tamman (VFT) behavior,

$$\omega_C = \omega_{C,0} \exp(-\frac{B}{T-T_{VFT}}) \tag{13}$$

where $\omega_{C,0}$ [s$^{-1}$] is constant, $T$ [K] is the direct current (DC) temperature, $B$ [K] and $T_{VFT}$ [K] are VFT parameters. By using the parameters summarized in Table 1, we calculate the temperature response at $r=0$, as shown in Fig. 2. One can find that the lower the DC temperature (except at 200 K), the longer the duration of the overshoot will be. Clearly, as indicated by the later setting in of the second shoulder, which is caused by the onset of channeling of vibrational energy to a configurational (volume increasing) change of the amorphous network, with decreasing DC temperature, the relaxation time of the specific heat capacity increases, as indicated by Eq.13. At 200K, the second shoulder is preceded by the thermal diffusion-driven washing away of the photothermally deposited heat and temperature gradient.

| Quantity | Values | Unit |
|---|---|---|
| $C_\infty$ | 1180 | J kg$^{-1}$ K$^{-1}$ |
| $C_0$ | 2100 | J kg$^{-1}$ K$^{-1}$ |
| $\omega_{C,0}$ | 5.75 10$^{14}$ | s$^{-1}$ |
| $B$ | 2210 | K |
| $T_{VFT}$ | 133 | K |
| $\kappa$ | 0.29 | W m$^{-1}$ K$^{-1}$ |
| $\rho$ | 1260 | Kg m$^{-3}$ |

Table 1 Thermal parameters of glycerol as determined from the literature [29,42]. These values are used for the simulation of Fig. 2.

## 2.2 Thermally induced displacement and density response in a relaxing system

In this part, we focus on the relaxation behavior of thermoelastic transients caused by an ultrashort laser pulse induced nonuniform temperature variation. We assume that the sample is described in the frame of the Kelvin-Voigt model for the viscoelasticity, corresponding with a lumped model containing a spring and a dashpot in parallel (as described on page 87 of Ref. [43]). Under this assumption, the constitutive equations are,

$$\begin{cases} \rho \frac{\partial^2 u}{\partial t^2} = \nabla . \sigma \\ \sigma = \varsigma \xi + \eta \frac{\partial \xi}{\partial t} \end{cases} \tag{14}$$

where $u$ [m] is the displacement, $\sigma$ [Pa] is the stress, $\varsigma$ [Pa] is the stiffness matrix, and $\eta$ [Pa s] is the viscosity. The strain $\xi$ can be expressed as,

$$\xi = \nabla_S u - \gamma_M \Delta T \tag{15}$$

where $\nabla_S u = (\nabla u + \nabla^T u)/2$, $\gamma_M$ is the matrix of linear expansion and $\Delta T$ is the temperature variation (taken from Eq. 12). This approach is in agreement with Green-Lindsay theory for thermoviscoelastic media [44,45]. Thus, in the case of an isotropic material, subject to a radial displacement $u = u_r(r, t)r$ (cylindrical coordinates), we can couple the two equations in Eq. 14 and Eq. 15 to get,

$$\left(1 + \frac{\eta}{\rho c_L^2}\frac{\partial}{\partial t}\right) c_L^2 \left(\frac{\partial^2 u_r}{\partial r^2} + \frac{1}{r}\frac{\partial u_r}{\partial r} - \frac{u_r}{r^2}\right) - \frac{\partial^2 u_r}{\partial t^2} = \left(3c_L^2 - 4c_T^2\right)\gamma \left(1 + \frac{\eta}{\rho c_L^2}\frac{\partial}{\partial t}\right)\frac{\partial \Delta T}{\partial r} \tag{16}$$

with $c_T$ [m s$^{-1}$] transverse bulk wave velocity, $c_L$ [m s$^{-1}$] longitudinal bulk wave velocity, and $\gamma$ [K$^{-1}$] the linear expansion coefficient. Applying Fourier transform to Eq. 16, we can get,

$$\left(1 + i\frac{\eta}{\rho c_L^2}\omega\right) c_L^2 \left(\frac{\partial^2 \tilde{u}_r}{\partial r^2} + \frac{1}{r}\frac{\partial \tilde{u}_r}{\partial r} - \frac{\tilde{u}_r}{r^2}\right) + \omega^2 \tilde{u}_r = \left(3c_L^2 - 4c_T^2\right)\left(1 + i\frac{\eta}{\rho c_L^2}\omega\right)\gamma \frac{\partial \Delta \tilde{T}}{\partial r} \tag{17}$$

By defining,

$$c^2(\omega) = c_L^2 + i\eta \frac{\omega}{\rho} \tag{18}$$

Eq. 17 can be re-written as,

$$\frac{\partial^2 \tilde{u}_r}{\partial r^2} + \frac{1}{r}\frac{\partial \tilde{u}_r}{\partial r} - \frac{\tilde{u}_r}{r^2} + \frac{\omega^2}{c^2(\omega)}\tilde{u}_r = \left(3 - 4\frac{c_T^2}{c_L^2}\right)\gamma \frac{\partial \Delta \tilde{T}}{\partial r} \tag{19}$$

Similarly, $\tilde{u}_r$ can also be expanded in terms of a Fourier Bessel series,

$$\tilde{u}_r(r,\omega) = \sum_{m=1}^{\infty} J_1(q_m r)\varphi_m(\omega) \tag{20}$$

where $q_m = j_{0m}/R$, $j_{1m}$ is the $m^{\text{th}}$ root of the first-order Bessel function $J_1(x)$, and $\varphi_m(\omega)$ is the corresponding Fourier Bessel coefficients. Inserting Eq. 20 into Eq. 19 and using the orthogonality properties of Bessel function (see Appendix C for a detailed calculation), the unknown Fourier Bessel coefficient $\varphi_m(\omega)$ can be solved as,

$$\varphi_m(\omega) = \frac{6 - 8\frac{c_T^2}{c_L^2}}{R^2 J_2^2(j_{1m})} \frac{\gamma c^2(\omega)}{\omega^2 - q_m^2 c^2(\omega)} \int_0^R \frac{\partial \tilde{T}}{\partial r} J_1(q_m r) r\, dr \tag{21}$$

From expression Eq. 9 we have,

$$\frac{\partial \Delta \tilde{T}}{\partial r} = \sum_{n=1}^{\infty} J_1(q_n r) \frac{Q_0 I_n q_n}{\pi \rho C_\infty J_1^2(j_{0n}) R^2} \times \frac{i(\omega - i\omega_C)}{(\omega - \omega_{1n})(\omega - \omega_{2n})} \quad (22)$$

By defining $F_{mn}(q_m, q_n) = \int_0^R J_1(q_n r) J_1(q_m r) r dr$, the displacement vector in the frequency domain can be expressed as,

$$\tilde{u}_r(r, \omega) = \sum_{m=1}^{\infty} \sum_{n=1}^{\infty} \left( \begin{array}{c} J_1(q_m r) \dfrac{Q_0 I_n q_n \left(6 - 8\dfrac{c_T^2}{c_L^2}\right)}{\pi \rho C_\infty R^4 J_1^2(j_{0n}) J_2^2(j_{1m})} F_{mn}(q_m, q_n) \\ \times \dfrac{i(\omega - i\omega_C)}{(\omega - \omega_{1n})(\omega - \omega_{2n})} \times \dfrac{\gamma c^2(\omega)}{\omega^2 - q_m^2 c^2(\omega)} \end{array} \right) \quad (23)$$

The linear thermal expansion coefficient has found to be an interesting relaxing physical quantity. Here, we assign again the Debye model to describe its relaxation behavior,

$$\gamma(\omega) = \gamma_\infty + \frac{\Delta \gamma}{1 + i\omega / \omega_\gamma} \quad (24)$$

where $\Delta \gamma = \gamma_0 - \gamma_\infty$, $\gamma_\infty$ is the high-frequency limit, and $\gamma_0$ is the low-frequency limit of the thermal expansion coefficients. $\omega_\gamma$ is characteristic (angular) frequency, which is temperature dependent. Upon substitution of the expression for $C^2(\omega)$ and $\gamma(\omega)$ into Eq. 19, we get the following expression of $\tilde{u}_r(r, \omega)$,

$$\tilde{u}_r(r, \omega) = -\sum_{m=1}^{\infty} \sum_{n=1}^{\infty} \left( \begin{array}{c} J_1(q_m r) \dfrac{Q_0 I_n q_n \gamma_\infty \eta \left(6 - 8\dfrac{c_T^2}{c_L^2}\right)}{\pi \rho^2 C_\infty R^4 J_1^2(j_{0n}) J_2^2(j_{1m})} F_{mn}(q_m, q_n) \\ \times \dfrac{(\omega - i\omega_C)(\omega - \omega_6)(\omega - \omega_7)}{(\omega - \omega_{1n})(\omega - \omega_{2n})(\omega - \omega_{3m})(\omega - \omega_{4m})(\omega - \omega_5)} \end{array} \right) \quad (25)$$

with,

$$\omega_{3m} = \frac{iq_m^2}{2\rho}(\eta - \sqrt{\eta^2 - \frac{4c_L^2\rho^2}{q_m^2}})$$

$$\omega_{4m} = \frac{iq_m^2}{2\rho}(\eta + \sqrt{\eta^2 - \frac{4c_L^2\rho^2}{q_m^2}})$$

$$\omega_5 = i\omega_\gamma \tag{26}$$

$$\omega_6 = i\omega_\gamma(1 + \frac{\Delta\gamma}{\gamma_\infty})$$

$$\omega_7 = i\rho\frac{c_L^2}{\eta}$$

Again, by applying the residue theorem, a transformation of Eq. 25 to time domain can be performed and returns,

$$u_r(r,t) = -\sum_{m=1}^{\infty}\sum_{n=1}^{\infty}\left\{J_1(q_m r)\frac{iQ_0 I_n q_n \gamma_\infty \eta\left(12 - 16\frac{c_T^2}{c_L^2}\right)}{\rho^2 C_\infty R^4 J_1^2(j_{0n})J_2^2(j_{1m})}F_{mn}(q_m,q_n) \times \left(\sum_{i=1}^{2}A_{in}\exp(i\omega_{in}t) + \sum_{j=3}^{4}A_{im}\exp(i\omega_{im}t) + A_5\exp(i\omega_5 t)\right)\right\}H(t) \tag{27}$$

with,

$$\begin{cases} A_{1n} = \dfrac{(\omega_{1n} - i\omega_C)(\omega_{1n} - \omega_6)(\omega_{1n} - \omega_7)}{(\omega_{1n} - \omega_{2n})(\omega_{1n} - \omega_{3m})(\omega_{1n} - \omega_{4m})(\omega_{1n} - \omega_5)} \\[6pt] A_{2n} = \dfrac{(\omega_{2n} - i\omega_C)(\omega_{2n} - \omega_6)(\omega_{2n} - \omega_7)}{(\omega_{2n} - \omega_{1n})(\omega_{2n} - \omega_{3m})(\omega_{2n} - \omega_{4m})(\omega_{2n} - \omega_5)} \\[6pt] A_{3m} = \dfrac{(\omega_{3m} - i\omega_C)(\omega_{3m} - \omega_6)(\omega_{3m} - \omega_7)}{(\omega_{3m} - \omega_{1n})(\omega_{3m} - \omega_{2n})(\omega_{3m} - \omega_{4m})(\omega_{3m} - \omega_5)} \\[6pt] A_{4m} = \dfrac{(\omega_{4m} - i\omega_C)(\omega_{4m} - \omega_6)(\omega_{4m} - \omega_7)}{(\omega_{4m} - \omega_{1n})(\omega_{4m} - \omega_{2n})(\omega_{4m} - \omega_{3m})(\omega_{4m} - \omega_5)} \\[6pt] A_5 = \dfrac{(\omega_5 - i\omega_C)(\omega_5 - \omega_6)(\omega_5 - \omega_7)}{(\omega_5 - \omega_{1n})(\omega_5 - \omega_{2n})(\omega_5 - \omega_{3m})(\omega_5 - \omega_{4m})} \end{cases} \tag{28}$$

From an experimental point of view, the most important is the relative density change or strain resulting from the changing temperature and pressure field, which is given as,

$$\varepsilon(r,t) = -\frac{\partial u_r(r,t)}{\partial r}$$

$$= \sum_{m=1}^{\infty}\sum_{n=1}^{\infty} \left( (J_0(q_m r) - J_2(q_m r)) \frac{iQ_0 I_n q_m q_n \gamma_\infty \eta \left(6 - 8\frac{c_T^2}{c_L^2}\right)}{\rho^2 C_\infty R^4 J_1^2(j_{0n}) J_2^2(j_{1m})} F_{mn}(q_m, q_n) \times \left( \sum_{j=1}^{2} A_{jn} \exp(i\omega_{jn} t) + \sum_{j=3}^{4} A_{jm} \exp(i\omega_{jm} t) + A_5 \exp(i\omega_5 t) \right) \right) H(t) \quad (29)$$

Here we give an analytical solution for the impulsive stimulated response for relative density or strain fluctuation, which gives intuitive insight into the mechanism behind transient pulse excitation. The first term of Eq. 29, which includes the damping parameters $\omega_{1n}$ and $\omega_{2n}$, are related to the thermal diffusion constant $q_n^2 \alpha_0$ and the relaxation parameters of specific heat capacity. The angular frequencies $\omega_{3m}$ and $\omega_{4m}$ are related to longitudinal bulk wave velocity and damping, so the second term represents damped sound wave contributions to the density response. The last term (include $\omega_5$) represents thermal expansion relaxation with relaxation time $1/\omega_\gamma$.

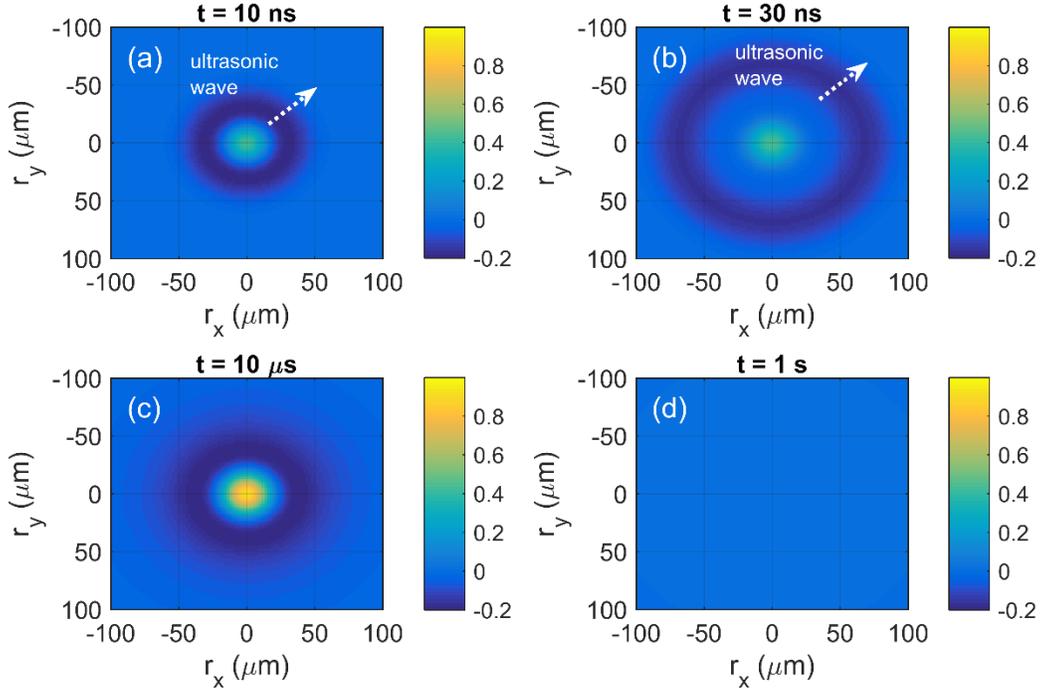

FIG. 3  Simulated evolution of the density response of glycerol at 230 K at the beam center ($z_1$) at several moments (a) t = 10 ns, (b) t = 30 ns, (c) t = 10 μs, and (d) t = 1s respectively. $r_x$ and $r_y$ represent the distance from the focus center of the pump laser $(0, 0, z_1)$ in $x$ and $y$ direction, seperately. The relationship between $r$ in Eq. 29 and $r_x$, $r_y$ satisfies $r = \sqrt{r_x^2 + r_y^2}$. The propagation of the laser induced acoustic pulses is marked with white arrows, which travels about 66 μm in 20 ns.

The same as $\omega_C$, assuming $\omega_\gamma$ also follows VFT-behavior,

$$\omega_\gamma = \omega_{\gamma,0} \exp(-\frac{B}{T-T_{VFT}}) \tag{30}$$

By using parameters summarized in Table 1 and Table 2, we simulated density response in the $x$ - $y$ plane at $z = z_1$ at different times. As illustrated in Fig. 3 (a-d), a clear TL effect can be observed after the deposition of the ultrashort laser pulse. This effect increases with the delayed volume expansion (Fig. 3 (a-c)), due to relaxation of heat capacity and thermal expansivity, and finally disappears on a longer time scale (Fig. 3 (d)) due to thermal diffusion. Accompanied by the local and transient thermal expansion, acoustic waves will inevitably be generated, which can also be seen in Fig. 3 (a) and Fig. 3 (b). Thus, the TL effect can be considered as the superposition of the temperature lens and the acoustic lens, thereby altering the propagating probe wavefront.

| Quantity | Values | Unit |
|---|---|---|
| $\gamma_\infty$ | $10^{-4}$ | $K^{-1}$ |
| $\gamma_0$ | $6\ 10^{-4}$ | $K^{-1}$ |
| $\omega_{\gamma,0}$ | $2.45\ 10^{14}$ | $s^{-1}$ |
| $c_L$ | 3300 | $m\ s^{-1}$ |
| $\eta$ | 1 | Pa s |

Table 2    Thermoelastic parameters of glycerol as determined from the literature [42,46]. These values are used in the simulation of Fig. 3.

**2.3 Phase shift and intensity calculation of the probe beam at the detector plane**

The TL detection is performed by analyzing the on-axis intensity variation of the central part of the probe beam in the far field, where the photodetector is located. The TL effect is based on this radially non-uniform optical phase delay, which acts on the light beam propagation in the same way as a lens. The phase shift can be expressed as,

$$\begin{aligned}\Delta\psi(r,t) &= k_p \Delta n L \\ &= k_p n \chi \varepsilon(r,t) L\end{aligned} \tag{31}$$

with $k_p = 2\pi/\lambda_p$ [m$^{-1}$] the wavenumber, $\lambda_p$ [nm] the wavelength of the probe beam. $L$ [mm] is the thickness of the sample and $\chi$ is the scale factor, which is a constant. The TL formed in the sample has a transmission function defined by $\exp(-i\Delta\psi(r, t))$. Thus, the TEM00 Gaussian probe beam in proximity of the sample can be expressed as [39],

$$U_p(r,z_1,t) = \sqrt{\frac{2P_p}{\pi}} \frac{1}{\omega_{1p}} \exp(-ik_p z_1) \times \exp\left[-i\left(\frac{k_p r^2}{2R_{1p}} + \Delta\psi(r,t)\right) - \frac{r^2}{\omega_{1p}^2}\right] \tag{32}$$

with $P_p$ [W] the total probe beam power, and $R_{1p}$ [m] the radius of curvature of the probe beam wavefronts at $z_1$. The probe beam propagating out of the sample to the detector plane can be obtained by applying Fresnel diffraction theory [47,48] to expression Eq. (32). As only the center point of the probe beam that passes through the pinhole is detected in our experiment, using cylindrical coordinates, Fresnel integration becomes [39],

$$U_p(z_1+z_2,t) = \frac{ik_p}{z_2}\exp(-ik_p z_2) \times \int_0^\infty U_p(r,z_1,t)\exp\left(-i\frac{k_p r^2}{2z_2}\right)rdr \qquad (33)$$

For the gaussian probe beam [39],

$$\omega_{1p}^2 = \omega_{0p}^2\left[1+(z_1/z_R)^2\right]$$
$$R_{1p} = (z_1^2+z_R^2)/z_1 \qquad (34)$$

where $z_R = \pi\omega_{0p}^2/\lambda_p$. Thus, inserting Eq. 32 into Eq. 33, we can get,

$$U_p(z_1+z_2,t) = A\int_0^\infty \exp(-i\Delta\psi(r,t))\exp\left(-(iV+1)\frac{r^2}{\omega_{1p}^2}\right)rdr \qquad (35)$$

where $A = \frac{ik_p}{z_2\omega_{1p}}\sqrt{\frac{2P_p}{\pi}}\exp(-ik_p(z_1+z_2))$, and $V = \frac{z_1}{z_R}+\frac{z_R}{z_2}\left(1+\frac{z_1^2}{z_R^2}\right)$. As the low optical absorption of the sample, the phase shift is very small ($\Delta\psi(r,t) \ll 1$). Thus, an approximation can be made $\exp(-i\Delta\psi(r,t)) \approx 1 - i\Delta\psi(r,t)$ [39]. Substituting Eq. 29 and Eq. 31 into Eq. 35, we can get,

$$U_p(z_1+z_2,t) \approx A\int_0^\infty (1-i\Delta\psi(r,t))\exp\left(-(iV+1)\frac{r^2}{\omega_{1p}^2}\right)rdr$$

$$= A\left\{\frac{\omega_{1p}^2}{2(1+iV)}+\sum_{m=1}^\infty\sum_{n=1}^\infty \left[\frac{\dfrac{Q_0 I_n q_m q_n \gamma_\infty \eta k_p n\chi L\left(6-8\dfrac{c_T^2}{c_L^2}\right)}{\rho^2 C_\infty R^4 J_1^2(j_{0n})J_2^2(j_{1m})}F_{mn}(q_m,q_n)}{\times\left(\sum_{i=1}^2 A_{in}\exp(i\omega_{in}t)+\sum_{j=3}^4 A_{im}\exp(i\omega_{im}t)\right)+A_5\exp(i\omega_5 t)}\right](F_0(q_m)-F_2(q_m))\right\}H(t)$$

(36)

with,

$$\begin{cases} F_0(q_m) = \int_0^\infty \exp\left(-(iV+1)\frac{r^2}{\omega_{1p}^2}\right) J_0(q_m r) r\, dr \\ F_2(q_m) = \int_0^\infty \exp\left(-(iV+1)\frac{r^2}{\omega_{1p}^2}\right) J_2(q_m r) r\, dr \end{cases} \quad (37)$$

where $F_0(q_m)$ and $F_2(q_m)$ are zero-order and second-order Hankel transform of $\exp\left(-(iV+1)\frac{r^2}{\omega_{1p}^2}\right)$, respectively. The final probe beam intensity [W m$^{-2}$] at the center of the detector plane can be expressed as,

$$I(z_1+z_2, t) = \frac{c_{light}\varepsilon_0}{2}\left|U_p(z_1+z_2, t)\right|^2 \quad (38)$$

with $c_{light}$ [m s$^{-1}$] the speed of light, and $\varepsilon_0$ [F m$^{-1}$] the permittivity of free space.

## 3. EXPERIMENT SETUP OF GLASSY DYNAMICS WITH TL

### 3.1 Experimental setup

Next, we will present an experimental investigation of the glassy dynamics in supercooled glycerol [49–52], using the time-resolved TL spectroscopy and the model developed above. Fig. 4 illustrates the experiment setup. A pulsed ND: YAG laser (Model Lab-130-10, Quanta-Ray®) operating at 1064 nm with a pulse width of 8 ns was used to excite the sample (> 99% glycerol in our experiment). The excitation laser was focused inside the sample using a 125 mm focal length lens (L). A continuous 532-nm TEM00 probe laser (Model Samba 100, Cobolt®), which merged coaxially through a dichroic mirror with the excitation laser, focused by the same lens, was used to probe the global density response to impulsive heating. The sample was sealed in a cuvette (optical path 2 mm, 45 mm (H) × 12 mm (L) × 12 mm (W)) and attached to the cold finger of an optical cryostat (Model Optistat-DN-V, Oxford Instruments®), allowing DC temperature control and/or scan over the sample.

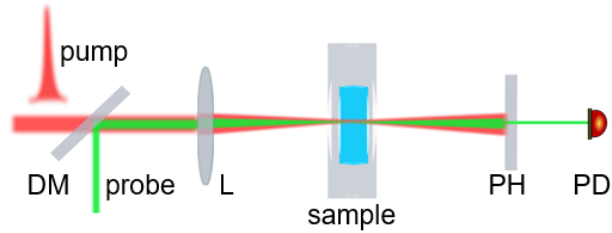

FIG. 4  Experimental setup of nanosecond laser-induced TL spectroscopy. DM, dichroic mirror; L, lens; PH, pinhole; PD, photodetector.

In this experiment, the focal waist of the probe laser (green) and pump laser (red) displaced with each other inside the sample as the aberration of the focusing lens [53], yielding the so-called mode-mismatched configuration [39]. After passing through the sample, the intensity of the probe beam was detected by a home-made photodetector (PD, bandwidth ~ 100 MHz) in the far field. A pinhole with a diameter of 1 mm was placed in front of the detector to enhance the detection of TL signals. Furthermore, an interference filter (IF) was placed in front of the pinhole to block the transmitted excitation light. The output of PD was coupled to a fast oscilloscope (LC564A, Lecroy®). In our experiment, the excitation laser was repeated at 10Hz, 500 TL cycles were recorded and averaged by the oscilloscope to reduce the noise. The experimental parameters of the system were summarized in Table 3.

| Quantity | Values | Unit |
|---|---|---|
| $\omega_e$ | 30 | μm |
| $\omega_{0p}$ | 100 | μm |
| $L$ | 2 | mm |
| $z_1$ | 3 | mm |
| $z_2$ | 1.5 | m |

Table 3      Experimental parameters of beam configuration for the geometry shown in Fig. 1.

### 3.2 Results and discussion

We recorded TL waveforms in a time window of about 20 ms at 12 selected temperatures from 280 K to 200 K. Fig. 5 summarizes all the waveforms (symbols) and the best fit (solid lines) with the model developed in this work, namely by Eq. 38[1]. Three subsequent processes can be observed separately: (1) a rapid increase in amplitude (0.5 ns to 80 ns), corresponding to the fast part of the temperature rise and resulting thermal expansion response to the sudden heat input; (2) a continued slow rise due to the slow part of the thermal expansion and underlying temperature dynamics; (3) the slow thermal diffusion from heat radially out of the heated laser beam path dominates, and the amplitude exponentially decays to zero, with a characteristic thermal diffusion decay time determined by the width of the pump laser beam at the focal point. With decreasing DC temperature, the second step occurs at later times, till it is quenched by the thermal diffusion decay of the signal. With increasing DC temperature, it occurs at earlier times, till it is overlapping with the initial, fast part of the thermal expansion. In a longer timescale, because more energy flowing to evoke cooperative rearrangements of the amorphous network, the heat capacity increases towards $C_0$, which dominated the thermal diffusion part. Worthy to mention is that, in a TL scheme, the measured global density response to impulsive heating can be considered as a convolution between the temperature response to impulsive heating (with the specific heat as response parameter) and the density response to a sudden temperature rise (with the thermal expansion coefficient as response parameter). Due to its time-varying and often spatially non-uniform character, the local thermal expansion response is unavoidably accompanied by the launching of acoustic waves, which carry information on the (relaxation behavior of the) elastic modulus. Besides, in many respects, the pulsed TL scheme is very similar to the one of the transient grating (TG), or impulsive stimulated thermal scattering (ISTS) [42,54]. The main difference between the two approaches lies in the geometry of the optical excitation pattern: while in TG the light pattern is spatially periodic and characterized by a single wavenumber, the Gaussian pattern used in a TL configuration results in a wide spectrum in the wavenumber domain. This difference in spectral content has mainly consequences concerning, 1) the thermal diffusion tail, which is purely exponential for TG signals and more complicated for TL signals; 2) acoustic signals at the beginning, which is a set of damped sinusoid oscillations for TG signals and a bipolar acoustic pulse in TL signals.

---

[1] For the sake of computation, the infinite sums over n amd m indexes present in Expression (38) have been truncated to $n = \bar{N}$ and m=$\bar{N}$. We have chosen $\bar{N} = 500$ as a good convergence value, the application of an higher $\bar{N}$ not altering the final result.

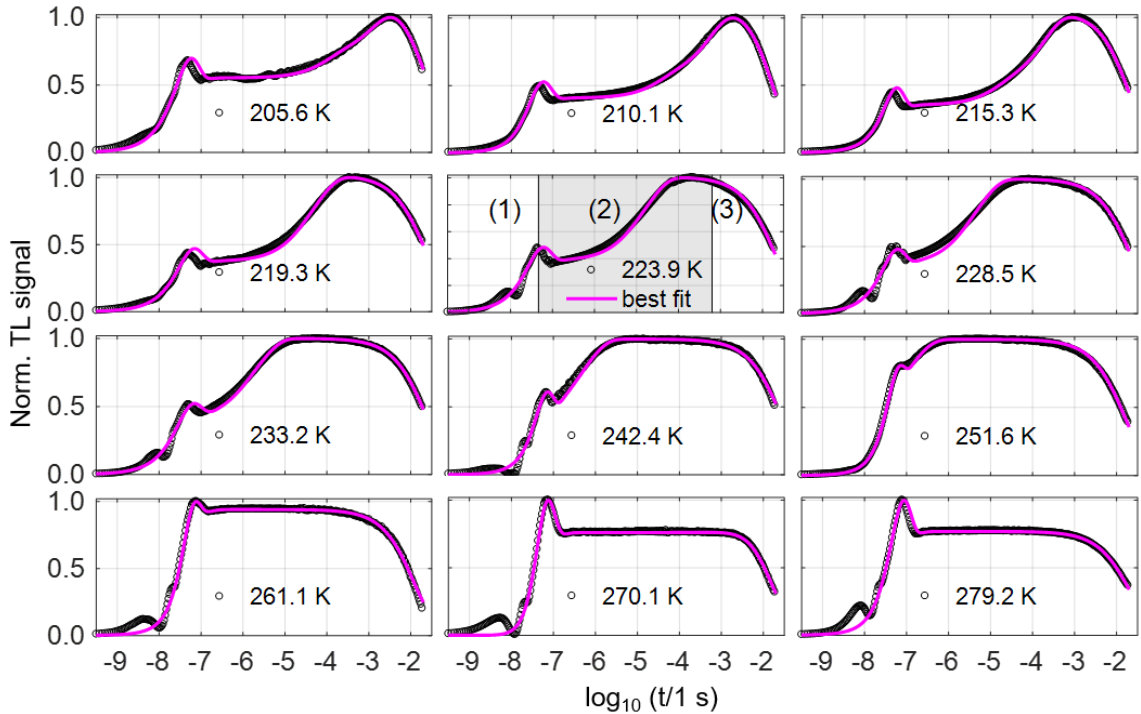

FIG. 5   Normalized TL signals of glycerol at a selection of temperatures (symbol, black) and best fits (solid line, magenta) by Eq. 38. A strongly (DC) temperature dependent double-step like density response can be observed at intermediate times (area (2)). The curves have been normalized to 1 at the maximum of the thermal diffusion value (maximum of the curve in area 3).

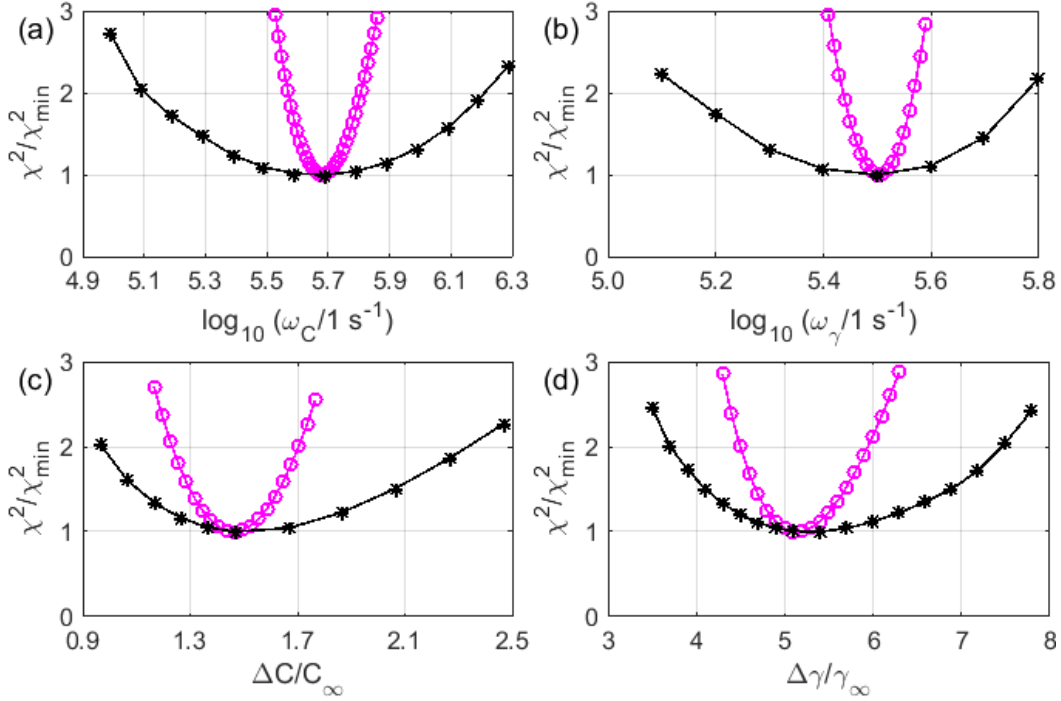

FIG. 6  Parabolic evolution of the least-squares (magenta circles) and most squares (black stars) cost function on the fitting parameters of (a) $\omega_\gamma$, (b) $\omega_C$, (c) $\Delta C/C_\infty$ and (d) $\Delta\gamma/\gamma_\infty$ for the TL signal.

Regarding the fitting, a protocol of most square error (MSE) analysis [55] was implemented to determine the fitting uncertainty. For each fitting parameter, the MSE analysis is done by evaluating the cost function, defined as the sum of the squared residuals (SSR) corresponding to the time vector $t$, Eq. 39, over a broad range centered around the best fitting values, $P_0$, while without fixing the rest fitting parameters. On the contrary, in a least-square error (LSE) analysis, the rest fitting parameters are fixed at their best fitting values, only the parameter to be evaluated is varied in a region its best fitting value.

$$\chi^2 = \frac{1}{N}\sum_{i=1}^{N}\left(TL_{\exp}(t_i) - TL_{fit}(t_i, P_0)\right)^2 \quad (39)$$

As an example, Fig. 6 illustrates MSE (circles) and LSE (squares) evaluation of the four best fit parameters for the fitting of the TL waveforms recorded at 233.2 K. Each curve shows a parabolical behavior of SSR around the best fitting value, suggesting nice convergence of the fitting/minimization procedure. The opening of the MSE curve is generally wider, as expected, than the LSE curve, since the former is able to takes into account the co-variance of the involved multiple fitting variable, namely, different combinations of fitting parameters yielding a statistically indistinguishable cost function value SSR (local minima). The finite width of the SSR parabola shows that the inverse problem of extracting the four fitting parameters from TL signal is feasible. Hence, TL spectroscopy and the model developed in this work allow to adequately determine thermal relaxation of specific heat capacity and thermal expansion coefficient.

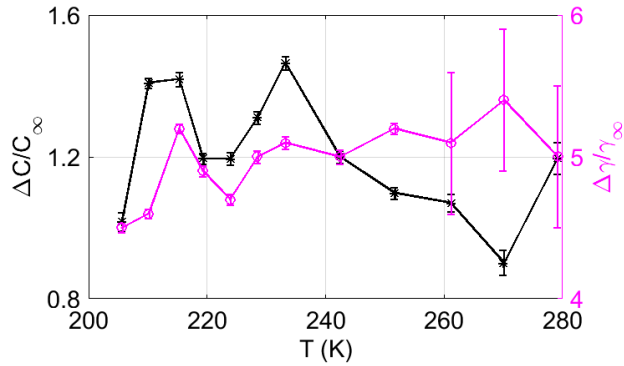

FIG. 7　　The fitting values of (left, black) $\Delta C/C_\infty$, and (right, magenta) $\Delta\gamma/\gamma_\infty$ versus DC temperature, error bar was obtained by most squares analysis.

Fig. 7 shows the fitted values of $\Delta C/C_\infty$ (left, black) and $\Delta\gamma/\gamma_\infty$ (right, magenta) versus temperature. Within the uncertainty margin, no temperature dependence is observed. $C_0$ is well defined in the thermal diffusion tail, and the average fitting value of $C_0$ is $2200 \pm 100$ J kg$^{-1}$ K$^{-1}$. We can then calculate $C_\infty$ with the ratio shown in Fig. 7, yielding $1020 \pm 70$ J kg$^{-1}$ K$^{-1}$, which consists well with results from PPE [29] and 3-$\omega$ [26,27]. In our fitting, $\gamma_\infty$ is fixed at $10^{-4}$ K$^{-1}$ [46], the average fitting value of $\gamma_0$ is $(6.0 \pm 0.3)\ 10^{-4}$ K$^{-1}$. Furthermore, the temperature dependence of the relative magnitude of the specific heat capacity relaxing and thermal expansion relaxation compared to their static values separately ($\Delta C/C_0$ and $\Delta\gamma/\gamma_0$), also termed as relaxation strength, can also be calculated, $0.54 \pm 0.04$ for specific heat capacity and $0.83 \pm 0.01$ for the thermal expansivity. Those values comply with literature values, 0.48 by 3-$\omega$ [26,27] and 0.44 by PPE [28,29] for specific heat capacity, 0.80 by DC dilatometry for expansivity [46].

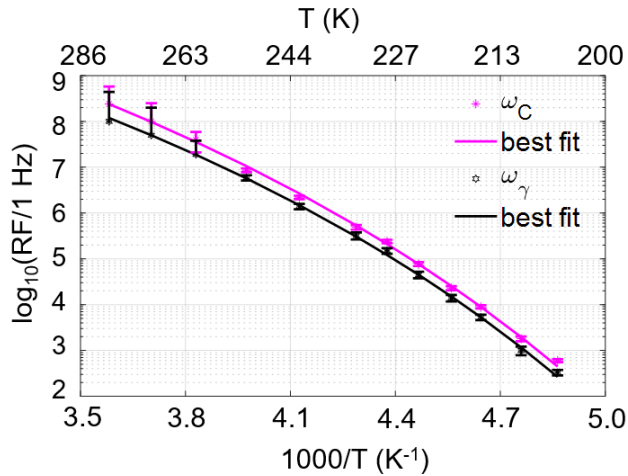

FIG. 8　　Arrhenius plot of the VFT behavior of the structural relaxation of glycerol determined by

TL spectroscopy (symbols) and the best fit (solid line). RF, relaxation frequency.

Figure 8 shows the fitted relaxation frequency of specific heat capacity ($\omega_C$) and thermal expansion ($\omega_\gamma$). The fitting uncertainties (error bar) are also obtained by most square analysis. Larger fitting uncertainties are found at high temperatures (> 260 K), when the relaxation frequency approaches 100 MHz This suggests the finite width of pump laser pulses, around 10 ns, has limited access to higher frequencies, which can be extended by using shorter laser pulses, e.g., sourced from picosecond or femtosecond lasers. The temperature dependence of relaxation frequency is fitted by the VFT equation (solid line), with $\omega_0$, $B$, and $T_{VFT}$ as fitting parameters. Table 4 summarized the VFT parameters for the two relaxing quantities investigated in this work by TL and their comparisons with the ones determined by other techniques, i.e. structural by ISTS [42], specific heat capacity by 3-$\omega$ [26,27] and PPE [28,29], the dielectric permittivity by broadband dielectric spectroscopy [14], and compliance by heterodyne ultrasonic spectrometer [56]. The fragility, which describes curvature of the Arrhenius plot, is defined by [57],

$$Fragility = 16 + \frac{590}{B/T_{VFT}} \quad (40)$$

For a given material between different response functions, the fragility is confirmed once more universal, despite the characteristic relaxation frequencies being somewhat different between different physical susceptibilities.

| Relaxation dynamics | Measurement technique | $\log_{10}(\omega_0/1\ Hz)$ | $B$ (K) | $T_{VFT}$ (K) | Fragility | Relaxation strength |
|---|---|---|---|---|---|---|
| Specific heat capacity | TL | 14.5 | 2147 | 127 | 50.9 | 0.54 |
| Thermal expansivity | TL | 13.9 | 2011 | 130 | 51.1 | 0.83 |
| Structural | ISTS | 14.7 | 2210 | 133 | 51.5 | 0.66 |
| Specific heat capacity | PPE | 11.9 | 1593 | 142 | 68.5 | 0.44 |
| Specific heat capacity | 3ω | 14.6 | 2500 | 128 | 46.2 | 0.48 |
| Compliance | Ultrasonic spectroscopy | 14.4 | 2310 | 129 | 48.9 | 0.6 |
| Dielectric | Dielectric spectroscopy | 14.0 | 2309 | 129 | 49.0 | - |

Table 4  Comparison of VFT behavior and relaxation strength of glycerol probed by thermal, mechanical, and dielectric susceptibilities.

## 4. CONCLUSIONS

In this paper, we have investigated the thermal relaxation dynamics in supercooled systems by making use of the ultrafast time-resolved Thermal Lens (TL) spectroscopy. Theoretically, we have developed analytically a model to describe the time-resolved TL response in a relaxing system by taking into account the relaxation of specific heat ($C$) and thermal expansivity ($\gamma$). Experimentally, we have presented a set of TL waveforms of supercooled glycerol in a broad time window, 1 ns-20 ms, in a wide temperature range,

200-280 K. The developed model has been used to fit the experimental waveforms, allowing the evaluation of several key relaxation features of $C$ and $\gamma$. The obtained low/high-frequency limit response, 2200±100 / 1020±70 J kg$^{-1}$ K$^{-1}$ for $C$ and $10^{-4}$ / 6.0±0.3 $10^{-4}$ K$^{-1}$ for $\gamma$, and the relaxation strength, 0.54±0.04 and 0.83±0.01, comply well with the ones determined by 3-$\omega$ [26,27], PPE [28,29], and DC dilatometry [46], confirming the reliability of TL model developed in this work and its correct application to study glassy dynamics. The approach has allowed to assess the slowing-down process of structural relaxation coupled with thermal dynamics, namely the VFT or non-Arrhenius behavior of the relaxation frequency from sub-kHz to sub-100 MHz, which largely extends the upper limit of the previously existing spectroscopy of $C$ and $\gamma$, 100 kHz and 1 Hz, achieved by PPE and capacitive scanning dilatometry [31–33]. It should be mentioned that the bandwidth can still be extended by making use of shorter laser pulses, 10 ns in this work, such as picosecond and femotosecond lasers. The obtained VFT plot of $C$ and $\gamma$ is parallel similar fragility, which is also comparable to that of dielectric and mechanical susceptibilities, confirming the universal relaxation behavior between the different response functions.

**APPENDIX A: DERIVATION OF THE EXPRESSION FOR Eq. 5**

The Bessel function of the first kind of order $v(v \in \mathbb{N})$ is defined by the convergent infinite series,

$$J_v(r) = \sum_{m=0}^{\infty} \frac{(-1)^m}{m!\Gamma(m+v+1)} (\frac{r}{2})^{2m+v}, (r \in \mathbb{R}) \tag{A1}$$

Thus,

$$\begin{aligned}
\frac{1}{r} J_1(q_n r) &= \frac{1}{r} \sum_{m=0}^{\infty} \frac{(-1)^m}{m!\Gamma(m+1+1)} (\frac{q_n r}{2})^{2m+1} \\
&= \frac{q_n}{2} \sum_{m=0}^{\infty} \frac{(-1)^m (m+1-m)}{m!\Gamma(m+1+1)} (\frac{q_n r}{2})^{2m} \\
&= \frac{q_n}{2} \left( \sum_{m=0}^{\infty} \frac{(-1)^m (m+1)}{m!\Gamma(m+1+1)} (\frac{q_n r}{2})^{2m} - \sum_{m=0}^{\infty} \frac{(-1)^m m}{m!\Gamma(m+1+1)} (\frac{q_n r}{2})^{2m} \right)
\end{aligned} \tag{A2}$$

As,

$$\sum_{m=0}^{\infty} \frac{(-1)^m (m+1)}{m!\Gamma(m+1+1)} (\frac{q_n r}{2})^{2m} = \sum_{m=0}^{\infty} \frac{(-1)^m}{m!\Gamma(m+0+1)} (\frac{q_n r}{2})^{2m+0} = J_0(q_n r) \tag{A3}$$

And,

$$\sum_{m=0}^{\infty}\frac{(-1)^m m}{m!\Gamma(m+1+1)}(\frac{q_n r}{2})^{2m} = \sum_{m=1}^{\infty}\frac{(-1)^m}{(m-1)!\Gamma(m+1+1)}(\frac{q_n r}{2})^{2m}$$

$$\overset{h=m-1}{=} \sum_{h=0}^{\infty}\frac{(-1)^{h+1}}{h!\Gamma(h+2+1)}(\frac{q_n r}{2})^{2h+2} \quad (A4)$$

$$= -J_2(q_n r)$$

Then we have,

$$\frac{1}{r}J_1(q_n r) = \frac{q_n}{2}\left(J_0(q_n r) + J_2(q_n r)\right) \quad (A5)$$

Inserting Eq. 4 into Eq. 3, and considering the expression in Eq. A5, we have,

$$\frac{1}{r}\frac{\partial}{\partial r}(r\frac{\partial \Delta \tilde{T}}{\partial r}) - \frac{i\omega}{\alpha}\Delta \tilde{T} = \sum_{n=1}^{\infty}\left(\frac{\partial^2 J_0(q_n r)}{\partial r^2} + \frac{1}{r}\frac{\partial J_0(q_n r)}{\partial r} - \frac{i\omega\rho C}{\kappa}J_0(q_n r)\right)\theta_n(\omega)$$

$$= \sum_{n=1}^{\infty}\left(-q_n^2\frac{J_0(q_n r) - J_2(q_n r)}{2} - q_n\frac{J_1(q_n r)}{r} - \frac{i\omega\rho C}{\kappa}J_0(q_n r)\right)\theta_n(\omega) \quad (A6)$$

$$= -\frac{1}{\kappa}\sum_{n=1}^{\infty}J_0(q_n r)\theta_n(\omega)(\kappa q_n^2 + i\omega\rho C)$$

Thus, we can finally get the expression of Eq. 5,

$$\sum_{n=1}^{\infty}J_0(q_n r)\theta_n(\omega)(\kappa q_n^2 + i\omega\rho C) = \frac{Q_0}{2\pi}\exp(-\frac{2r^2}{\omega_e^2}) \quad (A7)$$

**APPENDIX B: DERIVATION OF THE EXPRESSION FOR $\Delta T(r, t)$**

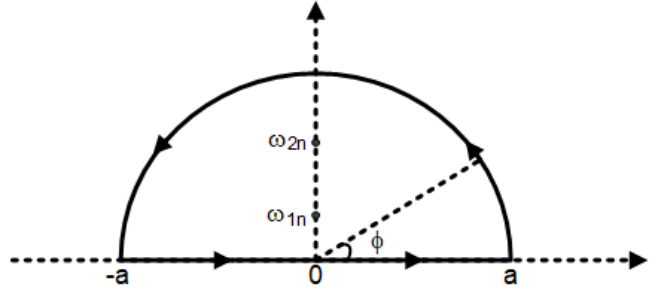

FIG. B    The contour $L$.

In this appendix we derive a detailed expression about how to get $\Delta T(r, t)$ by using residue theorem. In Eq. 9, we get the expression of $\Delta \tilde{T}(r, \omega)$. After inverse Fourier transform, we can get:

$$\Delta T(r,t) = \sum_{n=1}^{\infty}J_0(q_n r)\frac{Q_0 I_n}{\pi J_1^2(j_{0n})R^2} \times \int_{-\infty}^{\infty} f_n(\omega)\exp(i\omega t)d\omega \quad (B1)$$

with,

$$f_n(\omega) = \frac{-i(\omega - i\omega_C)}{\rho C_\infty (\omega - \omega_{1n})(\omega - \omega_{2n})} \tag{B2}$$

For $t > 0$, define the contour $L$ (as shown in Figure B) that goes along the real line from $-a \to a$ and then counterclockwise along a semicircle centered at 0 from $a \to -a$. For all $\varepsilon_n$, we have that $imag(\omega_{1n}) > 0$ and $imag(\omega_{2n}) > 0$. Taking $a$ to be greater than $abs(\omega_{1n})$ and $abs(\omega_{2n})$, so that the $\omega_{1n}$ and $\omega_{2n}$ are enclosed within the curve $L$. Thus, the function $f_n(\omega)$ has two singularities at $\omega = \omega_{1n}$ and $\omega = \omega_{2n}$, respectively. For the residue theorem, we have,

$$\oint_L f_n(\omega)\exp(i\omega t)\,d\omega = 2\pi i \sum_{j=1}^{2} Res(f_n(\omega)\exp(i\omega t), \omega_{jn}) \tag{B3}$$

The contour $L$ consists of a straight part and a curved arc ($arc = \{ae^{i\phi} | \phi \in [0, \pi]\}$). For the Jordan lemma theorem,

$$\left| \int_{arc} f_n(\omega)\exp(i\omega t)\,d\omega \right| \leq \frac{\pi}{t} max(|f_n(a\exp(i\phi))|) \tag{B4}$$

In our case, as $a \to \infty$, and,

$$\underset{a \to \infty}{limit}\left(max(|f_n(a\exp(i\phi))|)\right) \leq \underset{a \to \infty}{limit} \frac{\sqrt{a^2 + \omega_C^2}}{\sqrt{a^2 + \omega_{1n}^2}\sqrt{a^2 + \omega_{2n}^2}} = 0 \tag{B5}$$

Thus,

$$\int_{-\infty}^{\infty} f_n(\omega)\exp(i\omega t)\,dt = \int_{-\infty}^{\infty} f_n(\omega)\exp(i\omega t)\,dt + \underset{a \to \infty}{limit} \int_{arc} f_n(\omega)\exp(i\omega t)\,d\omega$$
$$= \underset{a \to \infty}{limit} \oint_L f_n(\omega)\exp(i\omega t)\,d\omega \tag{B6}$$

Finally, by using the residue theorem, we can obtain,

$$\Delta T(r,t) = \sum_{n=1}^{\infty} J_0(q_n r) \frac{Q_0 I_n}{\pi J_1^2(j_{0n})R^2} \times 2\pi i \sum_{i=1}^{2} Res(f_n(\omega)\exp(i\omega t), \omega_{in})$$
$$= \sum_{n=1}^{\infty} J_0(q_n r) \frac{2Q_0 I_n}{\rho C_\infty J_1^2(j_{0n})R^2} \times \left( \frac{\omega_{1n} - i\omega_C}{\omega_{1n} - \omega_{2n}} \exp(i\omega_{1n}t) + \frac{\omega_{2n} - i\omega_C}{\omega_{2n} - \omega_{1n}} \exp(i\omega_{2n}t) \right), (t > 0) \tag{B7}$$

For $t < 0$, define the contour $L'$ that goes along the real line from $a \to -a$ and then clockwise along a semicircle centered at 0 from $-a \to a$. For the Jordan lemma theorem, we can get,

$$\oint_{L'} f_n(\omega)\exp(i\omega t)\,d\omega = 0 \tag{B8}$$

Thus, we have,

$$\Delta T(r,t) = 0, (t < 0) \tag{B9}$$

Putting everything together, $\Delta T(r, t)$ can be solved as,

$$\Delta T(r,t) = \sum_{n=1}^{\infty} J_0(q_n r) \frac{Q_0 I_n}{\pi J_1^2(j_{0n})R^2} \times 2\pi i \sum_{i=1}^{2} Res(f_n(\omega)\exp(i\omega t), \omega_{in})$$

$$= \sum_{n=1}^{\infty} J_0(q_n r) \frac{2Q_0 I_n}{\rho C_\infty J_1^2(j_{0n})R^2} \times \left( \frac{\omega_{1n} - i\omega_C}{\omega_{1n} - \omega_{2n}} \exp(i\omega_{1n}t) + \frac{\omega_{2n} - i\omega_C}{\omega_{2n} - \omega_{1n}} \exp(i\omega_{2n}t) \right) H(t) \quad \text{(B10)}$$

## APPENDIX C: DERIVATION OF THE EXPRESSION FOR $\varphi_m(\omega)$

According to Eq. A5, we have,

$$\frac{1}{r^2} J_1(q_m r) = \frac{q_m}{2r}(J_0(q_m r) + J_2(q_m r)) \quad \text{(C1)}$$

Thus, considering the expression of $\tilde{u}_r$ in Eq. 20, we can get,

$$\frac{1}{r}\frac{\partial \tilde{u}_r}{\partial r} - \frac{\tilde{u}_r}{r^2} = \sum_{m=1}^{\infty} \left\{ \frac{q_m}{2r}(J_0(q_m r) - J_2(q_m r)) - \frac{q_m}{2r}(J_0(q_m r) + J_2(q_m r)) \right\} \varphi_m(\omega)$$

$$= -\sum_{m=1}^{\infty} \frac{q_m}{r} J_2(q_m r) \varphi_m(\omega) \quad \text{(C2)}$$

According to the definition of Bessel function in Eq. A1, we have,

$$\frac{q_m}{r} J_2(q_m r) = \frac{q_m}{r} \sum_{n=0}^{\infty} \frac{(-1)^n}{n!\Gamma(n+2+1)} \left( \frac{q_m r}{2} \right)^{2n+2}$$

$$= \frac{q_m^2}{4} \sum_{n=0}^{\infty} \frac{(-1)^n (n+2-n)}{n!\Gamma(n+2+1)} \left( \frac{q_m r}{2} \right)^{2n+1} \quad \text{(C3)}$$

$$= \frac{q_m^2}{4} \sum_{n=0}^{\infty} \frac{(-1)^n (n+2)}{n!\Gamma(n+2+1)} \left( \frac{q_m r}{2} \right)^{2n+1} - \frac{q_m^2}{4} \sum_{n=0}^{\infty} \frac{(-1)^n n}{n!\Gamma(n+2+1)} \left( \frac{q_m r}{2} \right)^{2n+1}$$

as,

$$\frac{q_m^2}{4} \sum_{n=0}^{\infty} \frac{(-1)^n (n+2)}{n!\Gamma(n+2+1)} \left( \frac{q_m r}{2} \right)^{2n+1} = \frac{q_m^2}{4} \sum_{n=0}^{\infty} \frac{(-1)^n}{n!\Gamma(n+1+1)} \left( \frac{q_m r}{2} \right)^{2n+1} = \frac{q_m^2 J_1(q_m r)}{4} \quad \text{(C4)}$$

and,

$$\frac{q_m^2}{4} \sum_{n=0}^{\infty} \frac{(-1)^n n}{n!\Gamma(n+2+1)} \left( \frac{q_m r}{2} \right)^{2n+1} = \frac{q_m^2}{4} \sum_{n=1}^{\infty} \frac{(-1)^n}{(n-1)!\Gamma(n+2+1)} \left( \frac{q_m r}{2} \right)^{2n+1}$$

$$\stackrel{h=n-1}{=} -\frac{q_m^2}{4} \sum_{h=0}^{\infty} \frac{(-1)^h}{h!\Gamma(h+3+1)} \left( \frac{q_m r}{2} \right)^{2h+3} \quad \text{(C5)}$$

$$= -\frac{q_m^2 J_3(q_m r)}{4}$$

Then we have,

$$\frac{q_m}{r} J_2(q_m r) = \frac{q_m^2 \left( J_1(q_m r) + J_3(q_m r) \right)}{4} \tag{C6}$$

Thus,

$$\frac{1}{r} \frac{\partial \tilde{u}_r}{\partial r} - \frac{\tilde{u}_r}{r^2} = -\sum_{m=1}^{\infty} \frac{q_m^2 \left( J_1(q_m r) + J_3(q_m r) \right)}{4} \varphi_m(\omega) \tag{C7}$$

Further, the second derivative of $\tilde{u}_r$ can be expressed as,

$$\frac{\partial^2 \tilde{u}_r}{\partial r^2} = \sum_{m=1}^{\infty} \frac{\partial^2 J_1(q_m r)}{\partial r^2} \varphi_m(\omega) = \sum_{m=1}^{\infty} \frac{q_m^2 \left( J_3(q_m r) - 3 J_1(q_m r) \right)}{4} \varphi_m(\omega) \tag{C8}$$

Then,

$$\frac{\partial^2 \tilde{u}_r}{\partial r^2} + \frac{1}{r} \frac{\partial \tilde{u}_r}{\partial r} - \frac{\tilde{u}_r}{r^2} = -\sum_{m=1}^{\infty} q_m^2 J_1(q_m r) \varphi_m(\omega) \tag{C9}$$

Inserting Eq. 20 into Eq. 19, we can get,

$$\sum_{m=1}^{\infty} \left( \frac{\omega^2}{c^2(\omega)} - q_m^2 \right) J_1(q_m r) \varphi_m(\omega) = \left( 3 - 4 \frac{c_T^2}{c_L^2} \right) \gamma \frac{\partial \Delta \tilde{T}}{\partial r} \tag{C10}$$

By using the orthogonality properties of Bessel function [41], multiplying $rJ_1(q_m r)$ in both sides of Eq. C10, and integrating over $r$ from 0 to $R$, we can get,

$$\varphi_m(\omega) = \frac{\left( 6 - 8 \frac{c_T^2}{c_L^2} \right)}{R^2 J_2^2(j_{1m})} \frac{\gamma c^2(\omega)}{\omega^2 - q_m^2 c^2(\omega)} \int_0^R \frac{\partial \tilde{T}}{\partial r} J_1(q_m r) r dr \tag{C11}$$

## ACKNOWLEDGEMENTS

CG and MG are grateful to the KU Leuven Research Council for financial support (C14/16/063 OPTIPROBE). MG acknowledges financial support by the National Research Council Joint Laboratories program, project SAC.AD002.026 (OMEN). LL acknowledges the financial support from FWO (Research Foundation-Flanders) postdoctoral research fellowship (12V4419N). PZ acknowledges the support of Chinese Scholarship Council (CSC). F. B. acknowledges financial support from Université de Lyon in the frame of the IDEXLYON Project (ANR-16-IDEX-0005) and from Université Claude Bernard Lyon 1 through the BQR Accueil EC 2019 grant.

## REFERENCES

[1]   C. A. Angell, *Formation of Glasses from Liquids and Biopolymers*, Science **267**, 1924 (1995).

[2]   M. D. Ediger, C. A. Angell, and S. R. Nagel, *Supercooled Liquids and Glasses*, J. Phys. Chem. **100**, 13200 (1996).
[3]   G. B. McKenna and S. L. Simon, *50th Anniversary Perspective: Challenges in the Dynamics and Kinetics of Glass-Forming Polymers*, Macromolecules **50**, 6333 (2017).
[4]   A. Safari, R. Saidur, F. A. Sulaiman, Y. Xu, and J. Dong, *A Review on Supercooling of Phase Change Materials in Thermal Energy Storage Systems*, Renew. Sustain. Energy Rev. **70**, 905 (2017).
[5]   G. G. Stonehouse and J. A. Evans, *The Use of Supercooling for Fresh Foods: A Review*, J. Food Eng. **148**, 74 (2015).
[6]   K. Kothari, V. Ragoonanan, and R. Suryanarayanan, *Influence of Molecular Mobility on the Physical Stability of Amorphous Pharmaceuticals in the Supercooled and Glassy States*, Mol. Pharm. **11**, 3048 (2014).
[7]   J. Wong and C. A. Angell, *Glass: Structure by Spectroscopy* (M. Dekker New York, 1976).
[8]   P. K. Dixon, L. Wu, S. R. Nagel, B. D. Williams, and J. P. Carini, *Scaling in the Relaxation of Supercooled Liquids*, Phys. Rev. Lett. **65**, 1108 (1990).
[9]   U. Bengtzelius, W. Gotze, and A. Sjolander, *Dynamics of Supercooled Liquids and the Glass Transition*, J. Phys. C Solid State Phys. **17**, 5915 (1984).
[10]  P. G. Debenedetti and F. H. Stillinger, *Supercooled Liquids and the Glass Transition*, Nature **410**, 259 (2001).
[11]  W. Gotze and L. Sjogren, *Relaxation Processes in Supercooled Liquids*, Rep. Prog. Phys. **55**, 241 (1992).
[12]  F. Sciortino, *Potential Energy Landscape Description of Supercooled Liquids and Glasses*, J. Stat. Mech. Theory Exp. **2005**, P05015 (2005).
[13]  J. C. Dyre, *Colloquium: The Glass Transition and Elastic Models of Glass-Forming Liquids*, Rev. Mod. Phys. **78**, 953 (2006).
[14]  P. Lunkenheimer, U. Schneider, R. Brand, and A. Loid, *Glassy Dynamics*, Contemp. Phys. **41**, 15 (2000).
[15]  T. Hecksher, D. H. Torchinsky, C. Klieber, J. A. Johnson, J. C. Dyre, and K. A. Nelson, *Toward Broadband Mechanical Spectroscopy*, Proc. Natl. Acad. Sci. **114**, 8710 (2017).
[16]  N. B. Olsen, T. Christensen, and J. C. Dyre, *Time-Temperature Superposition in Viscous Liquids*, Phys. Rev. Lett. **86**, 1271 (2001).
[17]  K. Niss and T. Hecksher, *Perspective: Searching for Simplicity Rather than Universality in Glass-Forming Liquids*, J. Chem. Phys. **149**, 230901 (2018).
[18]  T. C. Ransom and W. F. Oliver, *Glass Transition Temperature and Density Scaling in Cumene at Very High Pressure*, Phys. Rev. Lett. **119**, 025702 (2017).
[19]  W. Gotze and L. Sjogren, *Relaxation Processes in Supercooled Liquids*, Rep. Prog. Phys. **55**, 241 (1992).
[20]  W. Götze, *The Essentials of the Mode-Coupling Theory for Glassy Dynamics*, Condens. Matter Phys. (1998).
[21]  L. Janssen, *Mode-Coupling Theory of the Glass Transition: A Primer*, Front. Phys. **6**, 97 (2018).
[22]  J.-L. Garden, *Macroscopic Non-Equilibrium Thermodynamics in Dynamic Calorimetry*, Thermochim. Acta **452**, 85 (2007).
[23]  J.-L. Garden, *Macroscopic Non-Equilibrium Thermodynamics in Dynamic Calorimetry*, Thermochim. Acta **452**, 85 (2007).
[24]  D. Gundermann, U. R. Pedersen, T. Hecksher, N. P. Bailey, B. Jakobsen, T. Christensen, N. B. Olsen, T. B. Schrøder, D. Fragiadakis, and R. Casalini, *Predicting the Density-Scaling Exponent of a Glass-Forming Liquid from Prigogine–Defay Ratio Measurements*, Nat. Phys. **7**, 816 (2011).
[25]  N. O. Birge and S. R. Nagel, *Specific-Heat Spectroscopy of the Glass Transition*, Phys. Rev. Lett. **54**, 2674 (1985).

[26] N. O. Birge, P. K. Dixon, and N. Menon, *Specific Heat Spectroscopy: Origins, Status and Applications of the 3$ømega$ Method*, Thermochim. Acta **304**, 51 (1997).
[27] N. O. Birge, *Specific-Heat Spectroscopy of Glycerol and Propylene Glycol near the Glass Transition*, Phys. Rev. B **34**, 1631 (1986).
[28] E. H. Bentefour, C. Glorieux, M. Chirtoc, and J. Thoen, *Thermal Relaxation of Glycerol and Propylene Glycol Studied by Photothermal Spectroscopy*, J. Chem. Phys. **120**, 3726 (2004).
[29] E. H. Bentefour, C. Glorieux, M. Chirtoc, and J. Thoen, *Broadband Photopyroelectric Thermal Spectroscopy of a Supercooled Liquid near the Glass Transition*, J. Appl. Phys. **93**, 9610 (2003).
[30] Y. Z. Chua, G. Schulz, E. Shoifet, H. Huth, R. Zorn, J. W. Scmelzer, and C. Schick, *Glass Transition Cooperativity from Broad Band Heat Capacity Spectroscopy*, Colloid Polym. Sci. **292**, 1893 (2014).
[31] C. Bauer, R. Böhmer, S. Moreno-Flores, R. Richert, H. Sillescu, and D. Neher, *Capacitive Scanning Dilatometry and Frequency-Dependent Thermal Expansion of Polymer Films*, Phys. Rev. E **61**, 1755 (2000).
[32] K. Niss, D. Gundermann, T. Christensen, and J. C. Dyre, *Dynamic Thermal Expansivity of Liquids near the Glass Transition*, Phys. Rev. E **85**, 041501 (2012).
[33] K. Niss, D. Gundermann, T. Christensen, and J. C. Dyre, *Measuring the Dynamic Thermal Expansivity of Molecular Liquids near the Glass Transition*, ArXiv Prepr. ArXiv11034104 (2011).
[34] I. Chang, F. Fujara, B. Geil, G. Heuberger, T. Mangel, and H. Sillescu, *Translational and Rotational Molecular Motion in Supercooled Liquids Studied by NMR and Forced Rayleigh Scattering*, J. Non-Cryst. Solids **172–174**, 248 (1994).
[35] S. Sastry, P. G. Debenedetti, and F. H. Stillinger, *Signatures of Distinct Dynamical Regimes in the Energy Landscape of a Glass-Forming Liquid*, Nature **393**, 6685 (1998).
[36] B. Jakobsen, T. Hecksher, T. Christensen, N. B. Olsen, J. C. Dyre, and K. Niss, *Communication: Identical Temperature Dependence of the Time Scales of Several Linear-Response Functions of Two Glass-Forming Liquids* (American Institute of Physics, 2012).
[37] T. Kitamori, M. Tokeshi, A. Hibara, and K. Sato, *Peer Reviewed: Thermal Lens Microscopy and Microchip Chemistry* (ACS Publications, 2004).
[38] R. D. Snook and R. D. Lowe, *Thermal Lens Spectrometry. A Review*, Analyst **120**, 2051 (1995).
[39] J. Shen, R. D. Lowe, and R. D. Snook, *A Model for Cw Laser Induced Mode-Mismatched Dual-Beam Thermal Lens Spectrometry*, Chem. Phys. **165**, 385 (1992).
[40] J. Schroeder, *Signal Processing via Fourier-Bessel Series Expansion*, Digit. Signal Process. **3**, 112 (1993).
[41] S. K. Suslov, *Some Orthogonal Very Well Poised -Functions*, J. Phys. Math. Gen. **30**, 5877 (1997).
[42] D. M. Paolucci and K. A. Nelson, *Impulsive Stimulated Thermal Scattering Study of Structural Relaxation in Supercooled Glycerol*, J. Chem. Phys. **112**, 6725 (2000).
[43] B. A. Auld, *Acoustic Fields and Waves in Solids* (Рипол Классик, 1973).
[44] S. Mukhopadhyay, *Relaxation Effects on Thermally Induced Vibrations in a Generalized Thermoviscoelastic Medium with a Spherical Cavity*, J. Therm. Stress. **22**, 829 (1999).
[45] M. I. A. Othman and I. A. Abbas, *Fundamental Solution of Generalized Thermo-Viscoelasticity Using the Finite Element Method*, Comput. Math. Model. **23**, 158 (2012).
[46] I. V. Blaznov, N. P. Malomuzh, and S. V. Lishchuk, *Temperature Dependence of Density, Thermal Expansion Coefficient and Shear Viscosity of Supercooled Glycerol as a Reflection of Its Structure*, J. Chem. Phys. **121**, 6435 (2004).
[47] S. J. Sheldon, L. V. Knight, and J. M. Thorne, *Laser-Induced Thermal Lens Effect: A New Theoretical Model*, Appl. Opt. **21**, 1663 (1982).
[48] J. F. Power, *Pulsed Mode Thermal Lens Effect Detection in the near Field via Thermally Induced Probe Beam Spatial Phase Modulation: A Theory*, Appl. Opt. **29**, 52 (1990).

[49] B. Schiener, R. Böhmer, A. Loidl, and R. V. Chamberlin, *Nonresonant Spectral Hole Burning in the Slow Dielectric Response of Supercooled Liquids*, Science **274**, 752 (1996).

[50] P. Lunkenheimer, A. Pimenov, B. Schiener, R. Böhmer, and A. Loidl, *High-Frequency Dielectric Spectroscopy on Glycerol*, EPL Europhys. Lett. **33**, 611 (1996).

[51] L. Berthier, G. Biroli, J.-P. Bouchaud, L. Cipelletti, D. El Masri, D. L'Hôte, F. Ladieu, and M. Pierno, *Direct Experimental Evidence of a Growing Length Scale Accompanying the Glass Transition*, Science **310**, 1797 (2005).

[52] R. Richert, *Non-Linear Dielectric Signatures of Entropy Changes in Liquids Subject to Time Dependent Electric Fields*, J. Chem. Phys. **144**, 114501 (2016).

[53] The refractive index of the lens (N-BK7) is 1.5195 at 532 nm and 1.5066 at 1064 nm, respectively. Giving the curvature of the lens (LA1314, Thorlabs®) is 64.4 mm, the deviation of the waist position of the two beams could be estimated as ~ 3mm with the help of lens maker's equation, (n.d.).

[54] Gandolfi, Marco, *Impulsive Stimulated Scatttering Signal in Supercooled Liquids with Debye or Havriliak-Negami Relaxation of the Specific Heat Capacity and Thermal Expansion Coefficient*, in submission (n.d.).

[55] R. Salenbien, R. Cote, J. Goossens, P. Limaye, R. Labie, and C. Glorieux, *Laser-Based Surface Acoustic Wave Dispersion Spectroscopy for Extraction of Thicknesses, Depth, and Elastic Parameters of a Subsurface Layer: Feasibility Study on Intermetallic Layer Structure in Integrated Circuit Solder Joint*, J. Appl. Phys. **109**, 093104 (2011).

[56] Y. H. Jeong, S. R. Nagel, and S. Bhattacharya, *Ultrasonic Investigation of the Glass Transition in Glycerol*, Phys. Rev. A **34**, 602 (1986).

[57] R. Böhmer, K. L. Ngai, C. A. Angell, and D. J. Plazek, *Nonexponential Relaxations in Strong and Fragile Glass Formers*, J. Chem. Phys. **99**, 4201 (1993).